\documentclass[showpacs,aps,prd,reprint,superscriptaddress,nofootinbib]{revtex4-1}

\usepackage{graphicx}

\usepackage{amsmath,amssymb}
\usepackage{bm}

\usepackage{slashed}

\allowdisplaybreaks[1]

\usepackage[normalem]{ulem}  

\allowdisplaybreaks[3]

\newcommand\up{\uparrow}
\newcommand\down{\downarrow}

\begin{document}


\title{Kondo effect of $\bar{D}_{s}$ and $\bar{D}_{s}^{\ast}$ mesons in nuclear matter}


\author{Shigehiro~Yasui}
\email[]{yasuis@th.phys.titech.ac.jp}
\affiliation{Department of Physics, Tokyo Institute of Technology, Tokyo 152-8551, Japan}
\author{Kazutaka~Sudoh}
\affiliation{Nishogakusha University, 6-16, Sanbancho, Chiyoda, Tokyo 102-8336, Japan}



\begin{abstract}
We study the Kondo effect for $\bar{D}_{s}$ and $\bar{D}_{s}^{\ast}$ mesons as impurity particles in nuclear matter.
The spin-exchange interaction between the $\bar{D}_{s}$ or $\bar{D}_{s}^{\ast}$ meson and the nucleon induces the enhancement of the effective coupling in the low-energy scattering in the infrared region, whose scale of singularity is given by the Kondo scale.
We investigate the Kondo scale in the renormalization group equation at nucleon one-loop level.
We furthermore study the ground state with the Kondo effect in the mean-field approach, and present that the Kondo scale is related to the mixing strength between the $\bar{D}_{s}$ or $\bar{D}_{s}^{\ast}$ meson and the nucleon in nuclear matter.
We show the spectral function of the impurity when the Kondo effect occurs.
\end{abstract}

\pacs{12.39.Hg,14.40.Lb,14.40.Nd,21.85.+d}
\keywords{Charm/bottom nuclei, Kondo effect, Heavy quark effective theory}

\maketitle


\section{Introduction}

To study hadrons as impurity particles in nuclear medium is an important subject to investigate (i) the hadron-nucleon interaction, (ii) the change of nuclear medium by the impurity effect and (iii) the change of hadron properties from QCD vacuum to finite density.
There have been many studies for light flavor hadrons up to strangeness~\cite{Hayano:2008vn}.
Now the heavy flavor hadrons, i.e. charm and bottom, are investigated for future experiments at GSI-FAIR, J-PARC and so on~\cite{Friman:2011zz,Hosaka:2016ypm}.
We consider the {\it antiheavy-light} ($\bar{Q}q$) mesons, which are composed of a heavy anti-quark ($\bar{Q}=\bar{c},\bar{b}$) and a light quark ($q=u,d,s$), and investigate the {\it Kondo effect} induced by $\bar{Q}q$ mesons in nuclear mater.

The Kondo effect is the phenomena that the resistance of electrons in metals becomes enhanced at low temperature when there exist impurity atoms with finite spin~\cite{Hewson,Yamada}.
This is explained by (i) existence of degenerate state (Fermi surface), (ii) quantum effects and (iii) non-Abelian interaction, i.e. spin-exchange interaction, between the electron and the impurity atom, as long as the impurity is sufficiently heavy in mass, as discovered by Kondo~\cite{Kondo:1964}.
Due to the universality of the Kondo effect,
it is widely discussed not only in condensed matter but also in artificial materials and recently in nuclear and quark matter at extremely high density~\cite{Yasui:2013xr,Hattori:2015hka,Ozaki:2015sya,Yasui:2016ngy,Yasui:2016svc}.\footnote{See Ref.~\cite{Sugawara-Tanabe:1979} for early application of the Kondo effect to deformed nuclei.}

{\it Anticharm-strangeness} ($\bar{c}s$) mesons, i.e. $\bar{D}_{s}$ ($J^{P}=0^{-}$) and $\bar{D}_{s}^{\ast}$ ($1^{-}$) mesons with quark content $\bar{c}s$, are particularly interesting to study the Kondo effect in nuclear matter.
Due to spin 1 of the $\bar{D}_{s}^{\ast}$ meson, the spin-exchange interaction between the $\bar{D}_{s}^{\ast}$ meson and the nucleon provides the non-Abelian (spin-exchange) interaction relevant to the Kondo effect.
The $\bar{D}_{s}$ meson can also play the role in the spin-exchange interaction through the process $\bar{D}_{s}N \rightarrow \bar{D}_{s}^{\ast}N \rightarrow \bar{D}_{s}N$ for the nucleon $N$, although there is no direct spin-exchange between the $\bar{D}_{s}$ meson and the nucleon.
Because charm quark mass is much larger than the low-energy scale of the QCD, a few hundred MeV, we may regard approximately the charm quark mass infinitely large.
Then, we can adopt the heavy quark symmetry (HQS) as the fundamental spin symmetry based on QCD in the heavy quark mass limit~\cite{Neubert:1993mb,Manohar:2000dt}.
In terms of HQS, in fact, $\bar{D}_{s}$ and $\bar{D}_{s}^{\ast}$ mesons are considered to be the HQS partners.
Therefore, both $\bar{D}_{s}$ and $\bar{D}_{s}^{\ast}$ mesons should be included simultaneously as the dynamical degrees of freedom, leading to the mixing of $\bar{D}_{s}$ and $\bar{D}_{s}^{\ast}$ in nuclear matter.
In this case, both of them contribute to the Kondo effect.
The HQS must be better for  antibottom-strange mesons, i.e. $B_{s}$ and $B_{s}^{\ast}$ ($\bar{b}s$) mesons.
We denote $\bar{D}_{s}$ or $\bar{D}_{s}^{\ast}$ by $\bar{D}_{s}^{(\ast)}$ for a short notation.
In literature, there are many applications of the heavy quark symmetry to the anticharm(antibottom)-light flavor mesons in nuclear systems~\cite{Riska:1992qd,Oh:1994np,Oh:1994yv,Cohen:2005bx,Yasui:2009bz,Gamermann:2010zz,GarciaRecio:2011xt,Yamaguchi:2011xb,Yamaguchi:2011qw,Yamaguchi:2013hsa,Yasui:2012rw,Yasui:2013iga,Suenaga:2014dia,Suenaga:2014sga,Suenaga:2015daa} (see also references in Ref.~\cite{Hosaka:2016ypm}).

There are remarkable properties of $\bar{D}_{s}^{(\ast)}$ mesons, which are similar to or different from 
 $\bar{D}$ ($0^{-}$) and $\bar{D}^{\ast}$ ($1^{-}$) mesons with isospin 1/2.
\begin{enumerate}
\setlength{\itemsep}{0.0em}
\setlength{\leftskip}{1.2em}
\item There is no short-distance repulsion from the quark exchange in the interaction between a $\bar{D}_{s}^{(\ast)}$ meson and a nucleon. This is in contrast to the interaction between a $\bar{D}^{(\ast)}$ meson and a nucleon~\cite{Carames:2012bd}. Hence the $\bar{D}_{s}^{\ast}$ meson can access the deep inside of nuclear medium as $J/\psi$ in nuclear matter does~\cite{Brodsky:1989jd}.
\item The $\bar{D}_{s}^{(\ast)}$
 meson has an electric charge $-1$, hence it can be bound in atomic nuclei especially with large baryon numbers (e.g. $^{208}$Pb) thanks to the attraction from the Coulomb force even when the strong interaction is attractive only weakly.
The bound state may exist even for repulsion, if it is sufficiently weak. 
This is known also for the $\bar{D}^{-}$ meson in atomic nuclei~\cite{Tsushima:1998ru}. 
\item The $\bar{D}_{s}^{\ast}$ and $\bar{D}^{\ast}$ mesons can be regarded as a quasi-stable state in strong interaction in vacuum.
Although both of them can decay to a $\bar{D}_{s}$ or $\bar{D}$ meson via a pion emission, the decay width is very small due to the close thresholds~\cite{Agashe:2014kda}.
\item There is no isospin effect for the $\bar{D}_{s}^{(\ast)}$ meson due to isospin 0. This is complementary to the case of the $\bar{D}^{(\ast)}$ meson with isospin $1/2$, where both isospin-independent interaction and isospin-dependent interaction contribute simultaneously.
\end{enumerate}

By using these similarities and differences of $\bar{D}_{s}^{(\ast)}$ and $\bar{D}^{(\ast)}$ mesons, we can study the properties the Kondo effect for the $\bar{Q}q$ mesons in nuclear matter in multiple perspectives.
For example, it was shown that the Kondo effect induced by the isospin-exchange interaction for the $\bar{D}_{s}^{(\ast)}$ meson and the nucleon can exist in nuclear matter and in atomic nuclei~\cite{Yasui:2013xr,Yasui:2016ngy}.
This is complementary to the Kondo effect induced by the spin-exchange interaction.
We may consider the charge-conjugate state $D_{s}^{(\ast)}$ and $D^{(\ast)}$ mesons as well.
In this case, however, we have to consider additional channels such as $D_{s}N \rightarrow K\Lambda_{c}$, which should be covered elsewhere.
The difference of the properties between the $\bar{D}_{s}^{(\ast)}$ ($\bar{D}^{(\ast)}$) meson and the $D_{s}^{(\ast)}$ ($D^{(\ast)}$) meson arises essentially from the breaking of charge conjugation at finite baryon number density.
The study of $D_{s}^{(\ast)}$ and $D^{(\ast)}$ mesons is not covered in the present study.

The paper is organized as the following.
In Sec.~\ref{sec:interaction}, we introduce the interaction Lagrangian for the $\bar{D}_{s}^{(\ast)}$ meson and the nucleon based on HQS.
In Sec.~\ref{sec:Kondo_scale}, adopting the perturbative approach,
 we present that the effective interaction between a $\bar{D}_{s}^{(\ast)}$ meson and a nucleon in nuclear matter becomes enhanced at the low-energy scale in infrared region, whose scale of singularity is given by the Kondo scale.
Then, in Sec.~\ref{sec:mean_field}, we proceed to investigate the physical meaning of the Kondo scale beyond the perturbation.
We consider the mean-field approximation
 and show that the Kondo scale is in fact related to the mixing strength between the $\bar{D}_{s}^{(\ast)}$ meson and the nucleon, leading to the non-trivial behavior of the spectral function of the impurity particle in nuclear matter.
We find that HQS
 plays the significantly important role to realize the Kondo effect as the result of the mixing of the $\bar{D}_{s}$ meson and the $\bar{D}_{s}^{\ast}$ meson in nuclear matter.
The final section is devoted for the conclusion and the outlook.

\section{Interaction model}
\label{sec:interaction}
As the effective interaction between the nucleon ($\psi$) and the $\bar{D}_{s}^{(\ast)}$ meson ($P_{sv}^{(\ast)}$), we introduce the point-like (contact) interaction whose Lagrangian is given in general from chiral symmetry and HQS by
\begin{eqnarray}
 {\cal L}_{\mathrm{int}}
 =
 \frac{1}{2} \sum_{i} c_{i} \, \bar{\psi} \Gamma_{i} \psi \, \mathrm{Tr} \bar{H}_{sv} \Gamma_{i} H_{sv},
\label{eq:Lagrangian_NDs}
\end{eqnarray}
with the coupling constants $c_{i}$ ($i=1,\dots,5$) for the Dirac matrices $\Gamma_{1}=1$, $\Gamma_{2} = \gamma^{\mu}$, $\Gamma_{3}=\sigma^{\mu\nu}$, $\Gamma_{4}=\gamma^{\mu}\gamma_{5}$, $\Gamma_{5}=\gamma_{5}$.
The heavy-meson effective field
 is defined by~\cite{Neubert:1993mb,Manohar:2000dt}
\begin{eqnarray}
 H_{sv} = \left( \gamma^{\mu}P^{\ast}_{sv\mu} + i\gamma_{5}P_{sv} \right)\frac{1+v\hspace{-0.5em}/}{2},
\end{eqnarray}
in the frame with four-velocity $v^{\mu}$ for the vector-field $P^{\ast}_{sv\mu}$ for $(s\bar{Q})_{\mathrm{spin}\,1}$ ($v^{\mu}P^{\ast}_{sv\mu}=0$) and the pseudoscalar-field $P_{sv}$ for $(s\bar{Q})_{\mathrm{spin}\,0}$.
We define $\bar{H}_{sv} = \gamma^{0}H_{sv}^{\dag}\gamma^{0} \nonumber$.
We consider either proton or neutron for the nucleon, because $\bar{D}_{s}^{(\ast)}$ meson is blind to the nucleon isospin.

We consider the non-relativistic limit for the nucleon field and write $\psi=(\varphi,0)^{t}$ with two-component spinor $\varphi$.
By defining $c_{s} = -(c_{1}-c_{2})$ and $c_{t}=2c_{3}+c_{4}$,
we rewrite ${\cal L}_{\mathrm{int}}$ in the rest frame $v^{\mu}=(1,\vec{0})$ as
\begin{align}
{\cal L}_{\mathrm{int}}
=\,&
  c_{s} \varphi^{\dag}\!\varphi \! \left( \delta^{ij} P_{sv}^{\ast i \dag} P^{\ast j}_{sv} + P_{sv}^{\dag}P_{sv} \right) \nonumber \\
&\hspace{-1em}
 + ic_{t} \! \sum_{k} \! \varphi^{\dag} \! \sigma^{k} \! \varphi \!
 \left( \epsilon^{ijk} P_{sv}^{\ast i \dag} P^{\ast j}_{sv} \!-\! \left( P_{sv}^{\ast k\dag} P_{sv} \!-\! P_{sv}^{\dag}P^{\ast k}_{sv} \right) \right),
 \label{eq:DsN_Lagrangian}
\end{align}
with the Pauli matrices $\sigma^{k}$ ($k=1,2,3$) for spin. The first (second) term in the right-hand-side gives the spin-nonexchange (spin-exchange) interaction.

In terms of HQS, the $\bar{D}_{s}^{(\ast)}N$ state is classified to the HQS singlet state or the HQS doublet state~\cite{Yamaguchi:2014era,Yasui:2013vca}.
The HQS singlet channel is given by $-(1/2)\bar{D}_{s}N(^{2}S_{1/2})+(\sqrt{3}/2)\bar{D}_{s}^{\ast}N(^{2}S_{1/2})$ and the HQS doublet channel is given by $(\sqrt{3}/2)\bar{D}_{s}N(^{2}S_{1/2})+(1/2)\bar{D}_{s}^{\ast}N(^{2}S_{1/2})$ and $\bar{D}_{s}^{\ast}N(^{4}S_{3/2})$.
The interaction Lagrangian in Eq.~(\ref{eq:DsN_Lagrangian}) gives the couplings $-c_{s}-3c_{t}$ and $-c_{s}+c_{t}$ for the HQS singlet and doublet, respectively.
Therefore, the ground state for $c_{t}>0$ ($c_{t}<0$) in Eq.~(\ref{eq:DsN_Lagrangian}) should be the HQS singlet (doublet) state.
Notice that this is comparable with what is expected in the $\bar{D}^{(\ast)}N$ system.
The ground state of $\bar{D}^{(\ast)}N$ is the HQS doublet, when the one-pion-exchange potential is adopted~\cite{Yasui:2012rw,Yamaguchi:2011xb,Yamaguchi:2011qw,Yamaguchi:2013hsa}.

\section{Kondo scale}
\label{sec:Kondo_scale}

We consider the effective coupling for $c_{s}$ and $c_{t}$ in the interaction (\ref{eq:DsN_Lagrangian}) at low-energy scale in nuclear matter.
For this purpose, we apply the renormalization group equation in perturbation, assuming the small coupling constants.

We set the zero point of the heavy meson energy at the $\bar{D}_{s}^{\ast}$ meson mass.
This is a reasonable setting when the $\bar{D}_{s}^{\ast}$ meson is injected as a static particle with zero momentum into nuclear matter.
The mass position of the $\bar{D}_{s}$ meson is $-\delta M$.
We introduce the $P_{vs}$ and $P_{vs}^{\ast\mu}$ propagators with residual four-momentum $k^{\mu}$
\begin{align}
\frac{i}{2(v\!\cdot\!k+\delta M)+i\eta}, \hspace{1em}
\frac{i\delta_{ij}}{2v\!\cdot\!k+i\eta},
\end{align} 
for $P_{vs}$ and $P_{vs}^{\ast i}$ ($i=1,2,3$), respectively, with infinitesimally small and positive number $\eta>0$~\cite{Neubert:1993mb,Manohar:2000dt}.
The nucleon propagator with four-momentum $p^{\mu}$ is given by
\begin{align}
 \left( p\hspace{-0.44em}/ \!+\! m \right)
\left( 
 \frac{i}{p^{2}\!-\!m^{2}+i\eta} \!-\! 2\pi \theta(p_{0}) \delta(p^{2}\!-\!m^{2}) \theta(k_{\mathrm{F}}\!-\!|\vec{p}\,|)
\right),
\end{align}
with the nucleon mass $m$ and the Fermi momentum $k_{\mathrm{F}}$~\cite{Kaiser:2001jx}.

\begin{figure}[tb]
  \vspace*{-3em}
\includegraphics[scale=0.25]{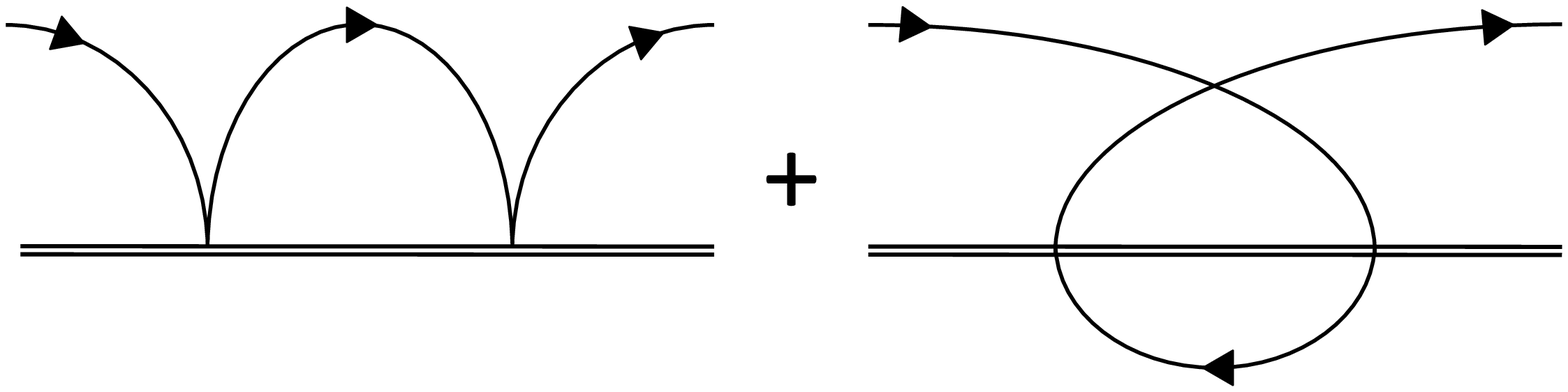}
  \vspace*{-7em}
 \caption{The diagrams at nucleon one-loop level. The single (double) solid line indicate the propagator of the nucleon ($\bar{D}_{s}^{(\ast)}$ meson).}
 \label{fig:diagram2}
\end{figure}

The coupling constants $c_{s}$ and $c_{t}$ in nuclear matter are not the same as those in vacuum, but get modified by the medium effect.
Following the ``poor man's scaling method"~\cite{0022-3719-3-12-008}, we consider the change of the coupling constants perturbatively by the renormalization group equation in nuclear matter as shown in Fig.~\ref{fig:diagram2}.
Considering the nucleon one-loop diagram, we obtain the renormalization group equations
\begin{eqnarray}
 \frac{\mathrm{d}}{\mathrm{d}\ell} c_{s00}(\ell) &=\,&  0, \label{eq:RGs00_2} \\
 \frac{\mathrm{d}}{\mathrm{d}\ell} c_{s11}(\ell) &=\,&  0, \label{eq:RGs11_2} \\
 \frac{\mathrm{d}}{\mathrm{d}\ell} c_{t10}(\ell) &=\,&   \frac{mk_{\mathrm{F}}}{2\pi^{2}} c_{t10}(\ell) c_{t11}(\ell), \label{eq:RGt10_2} \\
 \frac{\mathrm{d}}{\mathrm{d}\ell} c_{t01}(\ell) &=\,&  \frac{mk_{\mathrm{F}}}{2\pi^{2}} c_{t01}(\ell) c_{t11}(\ell), \label{eq:RGt10_2} \\
 \frac{\mathrm{d}}{\mathrm{d}\ell} c_{t11}(\ell) &=\,&  \frac{mk_{\mathrm{F}}}{2\pi^{2}} c_{t11}(\ell)^{2}, \label{eq:RGt11_2}
\end{eqnarray}
with $\ell=-\ln\Lambda/k_{\mathrm{F}}$ for the infrared momentum cutoff parameter $\Lambda$ below and above the Fermi surface in the loop integrals.
Here $c_{s00}$, $c_{s11}$, $c_{t11}$, $c_{t10}$ and $c_{t01}$ are the effective coupling constants for the vertices of $\varphi^{\dag}\varphi P_{sv}^{\dag}P_{sv}$, $\varphi^{\dag}\varphi P_{sv}^{\ast k \dag}P_{sv}^{\ast k}$, $\varphi^{\dag} \sigma^{k} \varphi \, \epsilon^{ijk} P_{sv}^{\ast i \dag}P_{sv}^{j}$, $\varphi^{\dag} \sigma^{k} \varphi P_{sv}^{\ast k \dag}P_{sv}$ and $\varphi^{\dag} \sigma^{k} \varphi P_{sv}^{\dag}P_{sv}^{\ast k}$ (summed over $i,j,k$) at scale $\Lambda$.
As the initial condition for the renormalization group equations, we consider $c_{s00}(0)=c_{s11}(0)=c_{s}$ and $c_{t11}(0)=c_{t10}(0)=c_{t01}(0) = c_{t}$ at the initial energy scale $\ell\simeq0$ ($\Lambda\simeq k_{\mathrm{F}}$).
As the solutions, we obtain $c_{s00}(\ell)=c_{s11}(\ell) = c_{s}$,
hence the effective coupling constants in $c_{s}$-term remain unchanged in nuclear matter.
In contrast, we obtain
\begin{eqnarray}
c_{t11}(\ell)=c_{t10}(\ell)=c_{t01}(\ell)=
\cfrac{c_{t}}{1-\cfrac{mk_{\mathrm{F}}}{2\pi^{2}}c_{t}\ell},
\label{eq:ct}
\end{eqnarray}
and hence the effective coupling constants in $c_{t}$-term changes drastically due to the singularity in the infrared energy scale for $c_{t}>0$, before the momentum cutoff parameter reaches the infrared limit $\ell \rightarrow \infty$ ($\Lambda \rightarrow 0$).
This scale of singularity is given by
\begin{eqnarray}
 \Lambda_{\mathrm{K}} = k_{\mathrm{F}} \exp \left( -\frac{2\pi^{2}}{mk_{\mathrm{F}}c_{t}} \right).
 \label{eq:Kondo_scale}
\end{eqnarray}
Therefore, the spin-exchange term ($c_{t}$-term) in Eq.~(\ref{eq:DsN_Lagrangian}) becomes logarithmically enhanced at low-energy scattering due to the loop effect for $c_{t}>0$.
The fixed point is given by $c_{t}^{\ast} \rightarrow \infty$ at $\Lambda \rightarrow \Lambda_{\mathrm{K}}$.
This is essentially the same as the Kondo effect known in the condensed matter physics~\cite{Hewson,Yamada}.
The energy scale $\Lambda_{\mathrm{K}}$ relevant for the Kondo effect is called the Kondo scale.
The Kondo effect does not break the heavy quark symmetry.

As a matter of course, for $c_{t}>0$, we should not take literary the singularity in Eq.~(\ref{eq:ct}).
This is rather a signal for the breakdown of perturbative treatment.
It indicates that the system of the $\bar{D}_{s}^{(\ast)}$ meson in nuclear matter becomes strongly interacting one in the low-energy scattering,
and it may lead to formation of non-perturbative objects such as bound and/or resonant states.
This problem will be discussed in the next section.

Notice that there is no singularity for $c_{t}<0$.
In this case, the fixed point is $c_{t}^{\ast} \rightarrow 0$ at $\ell \rightarrow \infty$ ($\Lambda \rightarrow 0$), and hence
 the perturbative treatment remains valid in the whole energy region due to the small coupling constant.
The physical meaning of this result is interesting.
First, the $\bar{D}_{s}^{\ast}$ meson does not have any spin-flip process in nuclear matter.
Second, the mixing between the $\bar{D}_{s}^{\ast}$ meson and the $\bar{D}_{s}$ meson in nuclear matter does not occur, because the $c_{t}$-term is only the mixing term.
Third, the $\bar{D}_{s}^{\ast}$ meson does not decay to the $\bar{D}_{s}$ meson, because the interaction process $\bar{D}_{s}^{\ast}N \rightarrow \bar{D}_{s}N$ for the nucleon $N$ vanishes.

We leave a comment before closing this section.
In the above calculation, it is important that the $\bar{D}_{s}^{\ast}$ meson mass is set to be at the Fermi surface.
If the $\bar{D}_{s}$ meson mass is at the Fermi surface, the scattering of the $\bar{D}_{s}^{(\ast)}$ meson and the nucleon is not affected by the infrared singularity, and the Kondo effect does not occur.
The choice of the energy zero point can be changed arbitrary.
However, it will be shown in the next section that the positions of the $\bar{D}_{s}^{(\ast)}$ meson masses will be determined uniquely in the mean-field approach.

\section{Mean-field approach}
\label{sec:mean_field}

\subsection{Hamiltonian with auxiliary fermion fields}

Let us investigate the ground state of the system under the Kondo effect for $c_{t}>0$.
For this purpose, we introduce the ``auxiliary" fermion fields.
The light-quark spin is decoupled from the heavy-quark spin in HQS, because the heavy-quark spin is independent of the interaction in the heavy quark limit~\cite{Neubert:1993mb,Manohar:2000dt}.
Hence, it is useful to replace the degrees of freedom from the $\bar{D}_{s}^{(\ast)}$ meson to the light quark ($s$ quark) in the $\bar{D}_{s}^{(\ast)}$ meson.
We call this light quark an auxiliary fermion, because this is no longer the free state but is confined in the finite region inside the $\bar{D}_{s}^{(\ast)}$ meson.

We denote the auxiliary fermion field by $\psi^{i}_{\sigma}$ which is labeled by the light-quark spin $\sigma=\up,\down$ and the heavy-quark spin $i=1,2$.
The labeling of the heavy-quark spin is introduced to specify the spin state of the heavy quark with which the auxiliary fermion is confined.\footnote{Here only the spin of the light quark is important, and the color is not necessary.}
Based on that $P_{sv}^{(\ast)}$ is given by the direct product of light-quark spin and heavy-quark spin and that the light-quark spin is expressed by $\psi^{i}_{\sigma}/2$,
we consider the correspondence between $P_{sv}^{(\ast)}$ and $\psi^{i}_{\sigma}$ given by
 $P_{sv} \leftrightarrow -( \psi^{2}_{\up} - \psi^{1}_{\down} )/2\sqrt{2}$,
 $P_{sv}^{\ast +1} \leftrightarrow \psi^{1}_{\up}/2$,
 $P_{sv}^{\ast 0} \leftrightarrow ( \psi^{2}_{\up} + \psi^{1}_{\down} )/2\sqrt{2}$ and
 $P_{sv}^{\ast -1} \leftrightarrow \psi^{2}_{\down}/2$
for  $P_{sv}^{\ast 1} = \left( -P_{sv}^{\ast +1} + P_{sv}^{\ast -1} \right)/\sqrt{2}$,
 $P_{sv}^{\ast 2} = i \left( -P_{sv}^{\ast +1} - P_{sv}^{\ast -1} \right)/\sqrt{2}$ and
 $P_{sv}^{\ast 3} = P_{sv}^{\ast 0}$.
The heavy-quark spin does not play any role in the dynamics, but only serves the labeling of $i$ for the light-quark spin.
In the following, the impurity means this auxiliary fermion instead of the $\bar{D}_{s}^{(\ast)}$ meson.
From Eq.~(\ref{eq:DsN_Lagrangian}), we consider only the $c_{t}$-term as the relevant term with the Kondo effect in the low-energy scattering, and give the interaction Hamiltonian
\begin{align}
 {\cal H}_{\mathrm{int}}
=
\frac{c_{t}}{4} \sum_{i,k} \varphi^{\dag} \sigma^{k} \varphi \, \psi^{i\dag} \sigma^{k} \psi^{i}.
\label{eq:int_H}
\end{align}

We change 
the interaction Hamiltonian
in the coordinate space to the one in the momentum space
by using the Fourier transformation
\begin{align}
 \varphi_{\sigma}(\bm{x}) =\,& \frac{1}{\sqrt{V}} \sum_{\bm{k}} e^{-i\bm{k}\cdot\bm{x}} \varphi_{\bm{k}\sigma},
 \label{eq:Fourier_varphi} \\
 \psi^{i}_{\sigma}(\bm{x}) =\,& \frac{1}{\sqrt{V}\sqrt{\sum_{\bm{k}'}}} \sum_{\bm{k}} e^{-i\bm{k}\cdot\bm{x}}
 \psi^{i}_{\sigma},
 \label{eq:Fourier_psi}
\end{align}
with the system volume $V$.
We introduce the normalization factor $1/\sqrt{\sum_{\bm{k}'}}$ for $\psi^{i}_{\sigma}$.
We assume that $\psi^{i}_{\sigma}$ is independent of the three-dimensional momentum $\bm{k}$ as the heavy impurity does not propagate in space.\footnote{We use the same notation $\psi^{i}_{\sigma}$ both in the coordinate space and in the momentum space.}
From Eq.~(\ref{eq:Fourier_psi}), we notice that the commutation relation $ \left\{ \psi^{i\dag}_{\sigma}, \psi^{j}_{\rho} \right\} = \delta^{ij} \delta_{\sigma\rho}$ is imposed from $\left\{ \psi^{i\dag}_{\sigma}(\bm{x}), \psi^{j}_{\rho}(\bm{y}) \right\} = \delta^{ij} \delta_{\sigma\rho} \delta^{(3)}(\bm{x}-\bm{y})$.
 
Importantly, because the auxiliary fermion ($\psi^{i}_{\sigma}$) is spatially confined in the $\bar{D}_{s}^{(\ast)}$ meson,
we introduce the constraint condition 
\begin{align}
 \sum_{\sigma} \psi^{i\dag}_{\sigma}(\bm{x}) \psi^{i}_{\sigma}(\bm{x}) = n^{i}\delta^{(3)}(\bm{x}),
  \label{eq:constraint_real}
\end{align}
with $n^{i}$ the probability of existence of the auxiliary fermion in the heavy-quark spin $i$ state, satisfying $n^{1}+n^{2}=1$.
The auxiliary fermion exists only at $\bm{x}=0$, where the location in the $\bar{D}_{s}^{(\ast)}$ meson is supposed to be the original point.
By using Eq.~(\ref{eq:Fourier_psi}), we find that the constraint condition (\ref{eq:constraint_real}) is expressed by
\begin{align}
 \sum_{\sigma} \psi^{i\dag}_{\sigma} \psi^{i}_{\sigma} = n^{i},
 \label{eq:constraint}
\end{align}
in the momentum space.

Considering the $c_{t}$-term, based on ${\cal H}_{int}$ (\ref{eq:int_H}), we obtain the Hamiltonian in momentum space
\begin{align}
 {H}_{\mathrm{eff}}
=\,&
\sum_{\bm{k},\sigma}
\tilde{\varepsilon}_{\bm{k}}
 \varphi_{\bm{k}\sigma}^{\dag} \varphi_{\bm{k}\sigma}
-\frac{C_{t}}{4} \!\!\! \sum_{\bm{k}_{1},\bm{k}_{2},\sigma} \!\!\!
  \varphi_{\bm{k}_{1}\sigma}^{\dag} \varphi_{\bm{k}_{2}\sigma}
\nonumber \\
&
+ 2 \frac{C_{t}}{4} \hspace{-1.5em} \sum_{\hspace{1em}i,\bm{k}_{1},\bm{k}_{2},\sigma_{1},\sigma_{2}} \hspace{-1.5em}
 \varphi_{\bm{k}_{1}\sigma_{1}}^{\dag} \varphi_{\bm{k}_{2}\sigma_{2}} \psi^{i\dag}_{\sigma_{2}} \psi^{i}_{\sigma_{1}}
\nonumber \\
&
+v \left( \psi^{1\dag}_{\down}-\psi^{2\dag}_{\up} \right)
\left( \psi^{1}_{\down}-\psi^{2}_{\up} \right)
+ \lambda \Bigl( \sum_{i,\sigma}  \psi^{i\dag}_{\sigma} \psi^{i}_{\sigma} -1 \Bigr),
\label{eq:H_eff}
\end{align}
with $\tilde{\varepsilon}_{\bm{k}}=\varepsilon_{\bm{k}} - \mu$ ($\mu \simeq k_{\mathrm{F}}^{2}/2m$) and $C_{t}=Vc_{t}$.
The term proportional to $v<0$ is for the HQS breaking, i.e. the mass splitting $\delta M=|2v|$ between $\bar{D}_{s}$ and $\bar{D}_{s}^{\ast}$ mesons.
The zero point of the meson energy is set to be the $\bar{D}_{s}^{\ast}$ meson mass.
To obtain Eq.~(\ref{eq:H_eff}), we use the constraint condition,
 $\sum_{i,\sigma} \psi^{i\dag}_{\sigma} \psi^{i}_{\sigma} = 1$,
from Eq.~(\ref{eq:constraint})
and add this condition in the last term with the Lagrange multiplier $\lambda$. 
We can consider the constraint condition
 in such a way  that the auxiliary fermion is confined on the hypersphere $S^{3}$ with unit radius.

\subsection{Mean-field approximation}
The mean-field approach has been known as a useful method to investigate the ground state with the Kondo effect in the condense matter physics~\cite{Takano:1966,Yoshimori:1970,Eto:2001} (see also Refs.~\cite{Hewson,Newns:1987}).
We apply the mean-field approximation for the four-point interaction of $\varphi^{\dag} \varphi \psi^{\dag} \psi$ in Eq.~(\ref{eq:H_eff}) as
\begin{align}
\hspace{-0.5em} \varphi_{\bm{k}\rho}^{\dag} \varphi_{\bm{l}\sigma} \psi^{i\dag}_{\sigma} \psi^{i}_{\rho}
\simeq&
-\langle \psi^{i\dag}_{\sigma} \varphi_{\bm{l}\sigma} \rangle
    \varphi_{\bm{k}\rho}^{\dag} \psi^{i}_{\rho}
   -\psi^{i\dag}_{\sigma} \varphi_{\bm{l}\sigma}
     \langle \varphi_{\bm{k}\rho}^{\dag} \psi^{i}_{\rho} \rangle
\nonumber \\
&
+\langle \psi^{i\dag}_{\sigma} \varphi_{\bm{l}\sigma} \rangle
  \langle \varphi_{\bm{k}\rho}^{\dag} \psi^{i}_{\rho} \rangle
+\delta_{\bm{k}\bm{\ell}} \delta_{\rho\sigma} \psi^{i\dag}_{\sigma} \psi^{i}_{\sigma}.
\end{align}
We define the ``gap" function
\begin{align}
 \Delta^{i}
\equiv
 -2\frac{C_{t}}{4} \sum_{\bm{k},\sigma}
 \langle \psi^{i\dag}_{\sigma} \varphi_{\bm{k}\sigma} \rangle,
 \label{eq:gap_definition}
\end{align}
as the spin-singlet condensate of the nucleon ($\varphi_{\bm{k}\sigma}$) and the auxiliary fermion ($\psi^{i}_{\sigma}$).
The gap $\Delta^{i}$ gives the mixing between the nucleon and the auxiliary fermion in the ground state.\footnote{In terms of HQS, the gap $(\Delta^{1},\Delta^{2})$ belongs to the HQS singlet, because the light degrees of freedom except for the heavy quark, i.e. the nucleon and the auxiliary fermion, form the spin-singlet.
}
This quantity is obtained by minimizing the total energy as the variational calculation, or by solving the self-consistent equation.\footnote{The positivity of $C_{t}>0$ is important to support the energy balance of the system with the condensate (\ref{eq:gap_definition}). This will be seen in the thermodynamic potential, Eqs.~(\ref{eq:thermo_pot_1}) and (\ref{eq:thermo_pot_2}). In negative case ($C_{t}<0$), we will have to consider spin-triplet condensate.}
In the following, as further approximation, we neglect the terms which are irrelevant to the direct coupling between the nucleon and the impurity, and consider the simplified Hamiltonian
\begin{align}
  {H}_{\mathrm{eff}}^{\mathrm{MF}\prime}
=\,&
\sum_{\bm{k},\sigma}
\tilde{\varepsilon}_{\bm{k}}
 \varphi_{\bm{k}\sigma}^{\dag} \varphi_{\bm{k}\sigma}
\nonumber \\
&\hspace{-1em}
+ \sum_{i}
 \biggl\{
 \sum_{\bm{k},\sigma}
 \left( \Delta^{i} \varphi_{\bm{k}\sigma}^{\dag} \psi^{i}_{\sigma} + \Delta^{i\dag} \psi^{i\dag}_{\sigma} \varphi_{\bm{k}\sigma} \right)
+
 \frac{|\Delta^{i}|^{2}}{2(C_{t}/4)}
 \biggr\}
\nonumber \\
&\hspace{-1em}
+v \left( \psi^{1\dag}_{\down}-\psi^{2\dag}_{\up} \right)
\left( \psi^{1}_{\down}-\psi^{2}_{\up} \right)
 + \lambda \Bigl( \sum_{i,\sigma}  \psi^{i\dag}_{\sigma} \psi^{i}_{\sigma} -1 \Bigr).
 \label{eq:H_eff_MF2}
\end{align}
This simplification does not change the essence of the discussion.

\subsection{Thermodynamic potential}

We discuss the gap by the variational calculation of the total energy.
For this purpose, we have to know the energy induced by the interaction between the nucleon and the impurity.

First of all, we consider the spectral function of the impurity in nuclear matter.
We define the Green's function ${\cal G}(\omega)$ by $(\omega+i\eta-\hat{h}){\cal G}(\omega)={\bf 1}$, where $\hat{h}$ is defined by ${H}_{\mathrm{eff}}^{\mathrm{MF}\prime} = \phi^{\dag} \hat{h} \phi + (|\Delta^{1}|^{2}+|\Delta^{2}|^{2})/2(C_{t}/4) - \lambda$ with $\phi = (\dots,\varphi_{\bm{k}\up},\dots,\psi^{1}_{\up},\psi^{2}_{\up},\dots,\varphi_{\bm{k}\down},\dots,\psi^{1}_{\down},\psi^{2}_{\down})^{t}$.
The spectral function relevant to the impurity is given by
\begin{align}
\bar{\rho}(\omega)
=\,&
-\frac{1}{\pi}\,\mathrm{Im}\,\mathrm{Tr}\, {\cal G}(\omega) - 2 \left( -\frac{1}{\pi} \mathrm{Im} \sum_{\bm{k}} \frac{1}{\omega+i\eta-\tilde{\varepsilon}_{\bm{k}}} \right)
\nonumber \\
=\,&
-\frac{1}{\pi} \, \mathrm{Im} \!\!\!
\sum_{a=1,2,3,4} \frac{1}{\omega+i\eta-\omega_{a}},
\label{eq:density_of_state}
\end{align}
with
\begin{align}
 \omega_{1} =\,& \frac{1}{2}(d^{1}+d^{2})+v-\sqrt{\frac{1}{4}(d^{1}+d^{2})^{2}+v^{2}}+\lambda, \\
 \omega_{2} =\,& \frac{1}{2}(d^{1}+d^{2})+v+\sqrt{\frac{1}{4}(d^{1}+d^{2})^{2}+v^{2}}+\lambda, \\
 \omega_{3} =\,& d^{1}+d^{2}+\lambda, \\
 \omega_{4} =\,& \lambda,
\end{align}
and
 $d^{i} = \sum_{\bm{k}} {|\Delta^{i}|^{2}}/{(\omega+i\eta-\tilde{\varepsilon}_{\bm{k}})}$
for $i=1,2$,
where the contribution from the free nucleon with spin factor two is subtracted.
In general, $\omega_{a}$ ($a=1,2,3,4$) is the complex number, $\omega_{a}=\omega_{a}^{R}+i\omega_{a}^{I}$, with the real part $(R)$ and the imaginary part $(I)$.
The spectral function is the sum of the Lorenzians with the pole position $\omega_{a}^{R}$ and the width $\omega_{a}^{I}$.

Using the spectral function (\ref{eq:density_of_state}), we obtain the thermodynamic potential (cf.~Refs.~\cite{Hewson,Eto:2001,Newns:1987})
\begin{align}
&{\Omega}(\lambda;\Delta^{1},\Delta^{2})
\nonumber \\
=\,&
-\frac{1}{\beta} \int_{-D}^{D} \ln (1+e^{-\beta\omega}) \bar{\rho}(\omega) \mathrm{d}\omega
+
\frac{|\Delta^{1}|^{2}+|\Delta^{2}|^{2}}{2(C_{t}/4)}
-\lambda,
\label{eq:thermo_pot_1}
\end{align}
where the cutoff scale $D$ ($\simeq k_{\mathrm{F}}^{2}/2m$) is introduced below and above the Fermi surface $(\omega=0)$ to regularize the integral.
At zero temperature ($\beta \rightarrow \infty$), regarding $D$ large, we approximately obtain
\begin{align}
& {\Omega}(\lambda;\Delta^{1},\Delta^{2})
\nonumber \\
\simeq&
\frac{1}{\pi}
\sum_{a}
\Biggl(
\omega_{a}^{I}
-
\omega_{a}^{R}
\arctan \frac{\omega_{a}^{I}}{\omega_{a}^{R}}
-
\frac{\omega_{a}^{I}}{2}
\ln \frac{{(\omega_{a}^{R}})^{2}+({\omega_{a}^{I}})^{2}}{D^{2}}
\Biggr)
\nonumber \\
&
+
\frac{|\Delta^{1}|^{2}+|\Delta^{2}|^{2}}{2(C_{t}/4)}
-\lambda.
\label{eq:thermo_pot_2}
\end{align}
We will consider only the imaginary part of $d^{i}$ for simplicity.
The values of $\Delta^{i}$ and $\lambda$ are obtained by the stationary conditions: ${\partial \Omega}/{\partial \Delta^{i}}=0$ and ${\partial \Omega}/{\partial \lambda}=0$.
In the following, we define $\delta^{i}=bV|\Delta^{i}|^{2}$ with $b=mk_{\mathrm{F}}/2\pi$ and $\delta=\delta^{1}+\delta^{2}$ for convenient notations.

We comment that it is a subtle problem which is the larger scale, the gap $\Delta^{i}$ (or $\delta$)
 or the HQS breaking scale $|v|$.
In the following, therefore, we consider the three case (i) $v=0$ (HQS limit), (ii) $|v|\ll\delta$ (small HQS breaking) and (iii) $|v|\gg\delta$ (large HQS breaking).

\begin{figure}[tb]
  \includegraphics[angle=-90, scale=0.23]{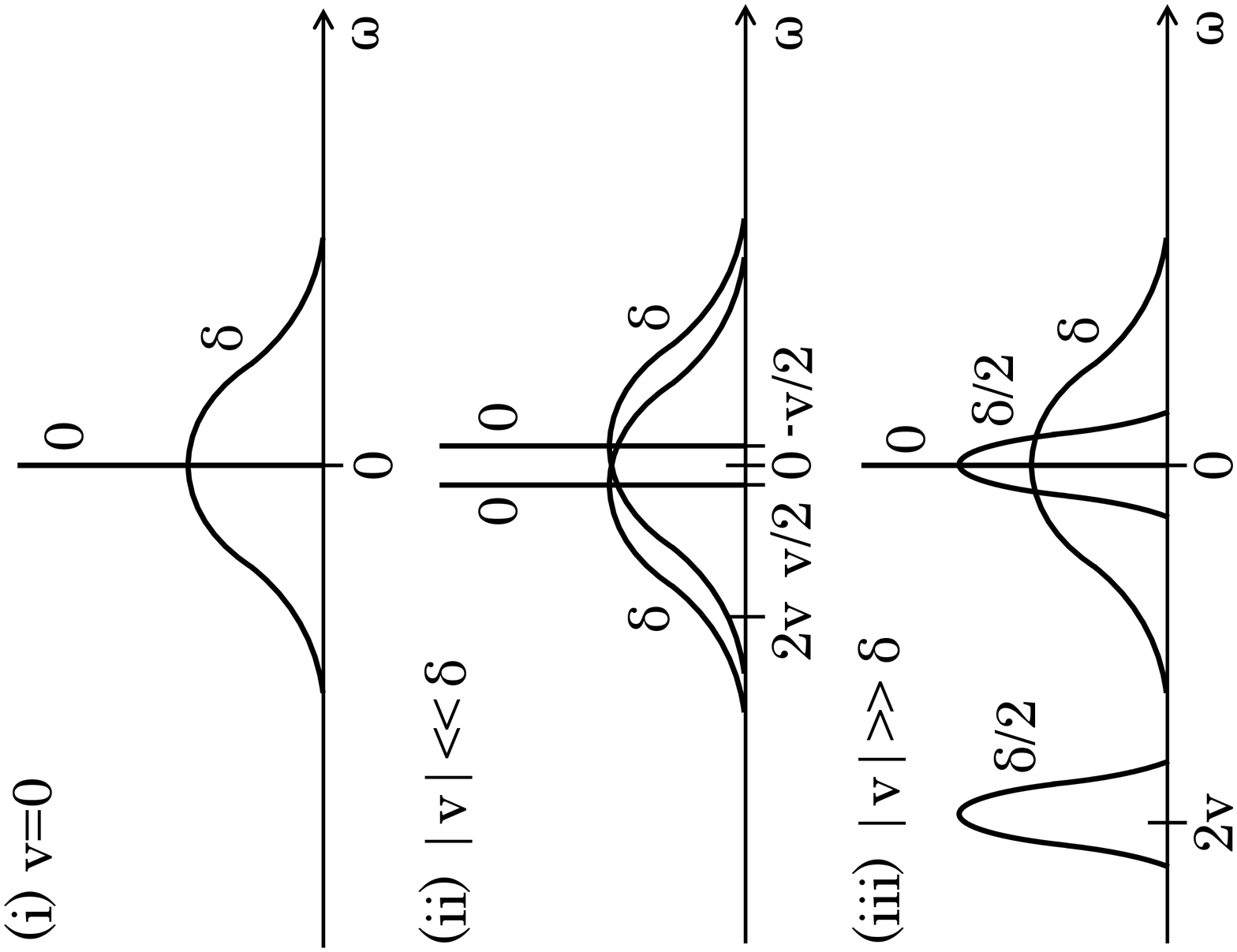}
 \caption{A schematic figure of the spectral function of impurities for (i) $v=0$, (ii) $|v| \ll \delta$ and (iii) $|v| \gg \delta$ for $\delta=\delta^{1}+\delta^{2}$. Widths are indicated in the figure. There are degeneracies in the delta-like peak and the broad peak, respectively, in (i).}
 \label{fig:160705}
\end{figure}

Before presenting the result, we show the sketch of the spectral function
(\ref{eq:density_of_state}) in Fig.~\ref{fig:160705}.
In the HQS limit (i), we have two delta-like peaks (width 0) and two broad resonances (width $\delta$) at $\omega=0$.
In the small HQS breaking (ii), we have a delta-like peak and a broad resonance  (width $\delta$) at $\omega=\mp v/2$ ($v<0$), respectively.
In the large HQS breaking (iii), we have a semi-broad peak (width $\delta/2$) at $\omega=2v$, and three peaks, i.e. a delta-like peak, a semi-broad peak (width $\delta/2$) and a broad peak (width $\delta$), at $\omega=0$.

In case (i), the formation of the mixing state between the nucleon ($\varphi_{\bm{k}\sigma}$) and the auxiliary fermion ($\psi^{i}_{\sigma}$) is realized by the superpositions of the heavy quark sectors $i=1,2$.
From Eq.~(\ref{eq:H_eff_MF2}), the combination of $\psi^{1}_{\sigma}$ and $\psi^{2}_{\sigma}$, i.e. $\propto \Delta^{1}\psi^{1}_{\sigma}+\Delta^{2}\psi^{2}_{\sigma}$, couples to the nucleon, while the orthogonal field, $\propto -\Delta^{2}\psi^{1}_{\sigma}+\Delta^{1}\psi^{2}_{\sigma}$, does not couple to the nucleon and stays free from the interaction.
The former and the latter modes correspond to the broad resonance and the delta-like peak, respectively.
In case (ii), the peak positions become split as the HQS breaking is added perturbatively.
In case (iii), it is reasonable that there is one state at $\omega=2v$ and three states at $\omega=0$, whose energy positions are in correspondence to the $\bar{D}_{s}$ and $\bar{D}_{s}^{\ast}$ meson masses.
Interestingly, the widths of the latter three peaks at $\omega=0$ are different.

In any case, it is important that the ``gap" $\Delta^{i}$ (or $\delta$) is related to the width of the peak, not to the position.
Thus, the gap is the quantity for the ``fluctuation" of the energy position of the impurity.
This can be understood by the definition of the gap $\Delta^{i}$ in Eq.~(\ref{eq:gap_definition}), because $\Delta^{i}$ gives the mixing between the nucleon and the auxiliary fermion in the ground state.
The resonant states with finite widths can be identified to the ``Kondo resonance" generated by the mixing between the nucleon and the impurity~\cite{Hewson}.

For the cases (i), (ii) and (iii), the gap and the thermodynamic potential can be given in the following.

\noindent{(i) $v=0$.}
We obtain
 $\lambda = 0$
and
\begin{align}
 \delta = \frac{1}{2}D \exp\!\left(-\frac{2\pi^{2}}{mk_{\mathrm{F}}c_{t}}\right).
 \label{eq:gap1}
\end{align}
The thermodynamic potential is
\begin{align}
 \Omega(\lambda;\Delta^{1},\Delta^{2}) = -\frac{2}{\pi}D \exp\!\left(-\frac{2\pi^{2}}{mk_{\mathrm{F}}c_{t}}\right).
\end{align}

\noindent{(ii) $|v|\ll\delta$.}
We calculate the quantities up to ${\cal O}(v)$, and obtain
 $\lambda \simeq -v/2$
and
\begin{align}
 \delta \simeq \frac{1}{2}D \exp\!\left(-\frac{2\pi^{2}}{mk_{\mathrm{F}}c_{t}}\right).
 \label{eq:gap2}
\end{align}
There is no additional term for the gap at ${\cal O}(v)$, but it appears at ${\cal O}(v^{2})$.
The thermodynamic potential is
\begin{align}
 \Omega(\lambda;\Delta^{1},\Delta^{2}) \simeq -\frac{2}{\pi}D \exp\!\left(-\frac{2\pi^{2}}{mk_{\mathrm{F}}c_{t}}\right) + \frac{v}{2}.
\end{align}

\noindent{(iii) $|v|\gg\delta$.}
We calculate only the leading order of large $|v|$, and obtain
 $\lambda \simeq 0$
and
\begin{align}
 \delta \simeq \frac{1}{2} \frac{D^{4/3}}{(-v)^{1/3}} \exp\!\left(-\frac{8\pi^{2}}{3mk_{\mathrm{F}}c_{t}}\right).
 \label{eq:gap3}
\end{align}
The thermodynamic potential is
\begin{align}
 \Omega(\lambda;\Delta^{1},\Delta^{2}) \simeq -\frac{3D^{4/3}}{2\pi(-v)^{1/3}} \exp\!\left(-\frac{8\pi^{2}}{3mk_{\mathrm{F}}c_{t}}\right).
 \label{eq:omega3}
\end{align}
In case (iii), the HQS breaking scale $|v|$ appears in a non-trivial manner in the gap as well as in the thermodynamic potential.
We remember that there is one semi-broad resonance at $\omega=2v$ in Fig.~\ref{fig:160705}(iii). 
Naively, we may expect that the Kondo resonance at $\omega=2v$ should vanish for the large $|2v|$ limit, because the interaction for the $\bar{D}_{s}$ meson and the nucleon $N$ is supplied by the process $\bar{D}_{s}N \rightarrow \bar{D}_{s}^{\ast}N \rightarrow \bar{D}_{s}N$ with an intermediate virtual $\bar{D}_{s}^{\ast}$ state, and this mixing should be suppressed for large $|v|$. 
This is indeed the case.
However, as long as the mass splitting $|2v|$ is kept to be large but finite,
there still exists the mixing between the $\bar{D}_{s}$ meson and the nucleon, provided the width as well as the binding energy  become smaller by the factor $(-D/v)^{1/3}$ and by the coefficient in the exponential as shown in Eqs.~(\ref{eq:gap3}) and (\ref{eq:omega3}), respectively.

In all the cases (i), (ii) and (iii), the nontrivial vacuum structure with a finite gap is realized as the ground state.
Such state is more stable than normal state because the thermodynamic potential is negative.
It is important also that the scale of this non-trivial vacuum is closely related to the Kondo scale $\Lambda_{\mathrm{K}}$ in Eq.~(\ref{eq:Kondo_scale}), because the $c_{t}$-dependence in $\delta$ and $\Omega$ for (i), (ii) and (iii) is expressed by powers of $\Lambda_{\mathrm{K}}$.
This suggests that the infrared singularity appeared in the perturbative analysis with the renormalization group equation is strongly related to the gap (\ref{eq:gap_definition}), i.e. the mixing between the nucleon and the auxiliary fermion, in the ground state.

So far, we have considered the $\bar{D}_{s}^{\ast}$ meson energy at the Fermi surface ($\omega=0$), and have obtained the spectral functions in Fig.~\ref{fig:160705}.
We may consider similarly the case that the $\bar{D}_{s}$ meson energy is at the Fermi surface by shifting the energy level upward uniformly by a constant value $-2v$.
Even in this case, we obtain the same spectral functions as in Fig.~\ref{fig:160705}(iii), namely a semi-broad peak at $\omega=2v$ and three different (delta-like, semi-broad, broad) peaks at $\omega=0$, provided that the total energy is shifted by the constant value $-2v$.
This results suggests that the realization of the overlapping between the energy position of the $\bar{D}_{s}^{\ast}$ meson and the Fermi surface is energetically most favored for the ground state.
In the mean-field approach, therefore, the energy position of the $\bar{D}_{s}^{\ast}$ meson is chosen automatically to minimize the total energy of the system with the Kondo effect.
This is consistent with the discussion in the perturbative approach.

Finally, we compare the present result with the ``triplet-singlet Kondo effect" in the condensed matter physics, which can be realized in quantum dot systems~\cite{Pustilnik:2000,Eto:2000,Eto:2001,Izumida:2001}.
It is known that the spin-triplet impurities in metal induces no Kondo effect, namely no enhancement of the effective coupling at the low-energy scattering, when higher order loops are included in the renormalization group equation (the ``underscreening" effect)~\cite{Nozieres:1980}.
However, the situation is different when the spin-singlet impurity exists in the energy region near the spin-triplet impurity.
In this case, the Kondo effect is induced by the mixing between the spin-triplet and spin-singlet impurities~\cite{Pustilnik:2000,Eto:2000,Eto:2001,Izumida:2001}.
This is analogous to the Kondo effect for the $\bar{D}_{s}^{\ast}$ and $\bar{D}_{s}$ mesons in nuclear matter in the present study.
In our case, the $\bar{D}_{s}^{\ast}$ ($\bar{D}_{s}$) meson is assigned to the spin-triplet (spin-singlet) impurity.
As a starting point, we have considered the mixing between the $\bar{D}_{s}^{\ast}$ meson and the $\bar{D}_{s}$ meson as a natural result of introducing HQS in Eq.~(\ref{eq:DsN_Lagrangian}).
When we included only the $\bar{D}_{s}^{\ast}$ meson in the renormalization group with higher nucleon loops beyond the one-loop level in Sec.~\ref{sec:Kondo_scale}, we may expect that there could be no Kondo effect due to the ``underscreening" effect.
When the $\bar{D}_{s}$ meson also is included, however, the mixing between the $\bar{D}_{s}^{\ast}$ meson and the $\bar{D}_{s}$ meson should induce the Kondo effect by the ``triplet-singlet" mixing mechanism.
In fact, this is found in the mean-field approach as presented above.

\section{Conclusion and outlook}
We discuss the $\bar{D}_{s}^{(\ast)}$ meson in nuclear matter.
Considering the spin-exchange interaction between the $\bar{D}_{s}^{(\ast)}$ meson and the nucleon,
we find that the effective coupling constant becomes logarithmically enhanced due to the Kondo effect at the low-energy scattering for $c_{t}>0$, whose scale of singularity is given by the Kondo scale.
For $c_{t}<0$, in contrast, the mixing between the $\bar{D}_{s}^{\ast}$ meson and the $\bar{D}_{s}$ vanishes in the low-energy scattering, and the $\bar{D}_{s}^{\ast}$ meson remains as a stable state in nuclear matter.
As for the enhanced effective coupling for $c_{t}>0$, we furthermore analyze the ground state in the mean-field approach,
and find that the Kondo scale is related to the mixing strength between the $\bar{D}_{s}^{(\ast)}$ meson and the nucleon, leading to the non-trivial behaviors of the spectral function of the impurity.
The spectral function of the impurity is the important information for experimental research of the Kondo effect of the $\bar{D}_{s}^{(\ast)}$ meson in nuclear matter.
The present result can be straightforwardly applied to the $B_{s}^{(\ast)}$ meson.

We have considered either proton or neutron as the nucleon isospin.
When we consider both proton and neutron simultaneously,
we will find easily that the coherent state as a sum of the proton and the neutron contributes to the coupling to the $\bar{D}_{s}^{(\ast)}$ meson, and the other combination orthogonal to the coherent state remains as a free state with no coupling to the $\bar{D}_{s}^{(\ast)}$ meson.
The spectral function of the $\bar{D}_{s}^{(\ast)}$ meson is the same.

There are several unsolved problems.
In the present analysis, we consider only $\bar{D}_{s}^{(\ast)}N$ interaction and neglected $\bar{D}^{(\ast)}Y$ ($Y=\Lambda,\Sigma$), because the latter channel is not directly related to the Kondo effect.
However, for more precise study, considering the latter channel must be important.
For example, it is known that there can exist a strong interaction in the $\bar{D}_{s}N$-$\bar{D}\Lambda$-$\bar{D}\Sigma$ channel so that the resonant state is formed at mass position about 2780 MeV~\cite{Hofmann:2005sw}, corresponding to the exotic $\bar{c}sqqq$ state discussed in the quark model~\cite{Lipkin:1987sk}.
Application of the Kondo effect to other heavy-light mesons such as $\bar{D}^{(\ast)}$ mesons as well as $D_{s}^{(\ast)}$ and $D^{(\ast)}$ mesons \cite{Oh:1994ux,Oh:1995ey,GarciaRecio:2008dp,GarciaRecio:2012db,Yamaguchi:2013ty,Tolos:2009nn,GarciaRecio:2010vt} will be an important subject.
The competition or cooperation between the isospin-exchange interaction and the spin-exchange interaction will be interesting.
Studying the chiral partners
 is important also.
Particularly, chiral partners of the $\bar{D}_{s}^{(\ast)}$ meson, the $\bar{D}_{s0}^{\ast}(2317)$ $(0^{+})$ meson and the $\bar{D}_{s1}(2460)$ $(1^{+})$ meson, are quasi-stable states with small decay widths,
and hence they will be useful
 for investigating the partial restoration of the broken chiral symmetry in medium~\cite{Sasaki:2014asa,Sasaki:2014wma}.
Heavy-light-light ($Qqq$) baryons~\cite{Liu:2011xc,Maeda:2015hxa} and heavy-heavy-light ($QQq$) baryons as well as heavy-heavy meson ($\bar{Q}Q$) are also interesting objects to be investigated for the Kondo effect.
Those subjects are left for future works.

\section*{Acknowledgments}
The authors thank M.~Oka for fruitful discussions and careful reading of draft, and thank A.~Yokota and K.~Ohtani for useful comments. This work is supported by the Grant-in-Aid for Scientific Research (Grant No.~25247036 and No.~15K17641) from Japan Society for the Promotion of Science (JSPS).

\end{document}